# Obfuscated Memory Malware Detection

Sharmila S P [1,2], Aruna Tiwari [1], Narendra S Chaudhari [1]

[1] Computer Science and Engineering, Indian Institute of Technology Indore, Madhya Pradesh, India
{phd2201101012, artiwari, nsc}@iiti.ac.in

[2] Information Science and Engineering, Siddaganga Institute of Technology, Tumakuru, Karnataka, India
{sharmila@sit.ac.in}

*Abstract*—Providing security for information is highly critical in the current era with devices enabled with smart technology, where assuming a day without the internet is highly impossible. Fast internet at a cheaper price, not only made communication easy for legitimate users but also for cybercriminals to induce attacks in various dimensions to breach privacy and security. Cybercriminals gain illegal access and breach the privacy of users to harm them in multiple ways. Malware is one such tool used by hackers to execute their malicious intent. Development in AI technology is utilized by malware developers to cause social harm. In this work, we intend to show how Artificial Intelligence and Machine learning can be used to detect and mitigate these cyber-attacks induced by malware in specific obfuscated malware. We conducted experiments with memory feature engineering on memory analysis of malware samples. Binary classification can identify whether a given sample is malware or not, but identifying the type of malware will only guide what next step to be taken for that malware, to stop it from proceeding with its further action. Hence, we propose a multi-class classification model to detect the three types of obfuscated malware with an accuracy of 89.07% using the Classic Random Forest algorithm. To the best of our knowledge, there is very little amount of work done in classifying multiple obfuscated malware by a single model. We also compared our model with a few state-of-the-art models and found it comparatively better.

## I. INTRODUCTION (*HEADING 1*)

Rapid development in digital technology and the replacement of paper documents with e-documents have caused a hike in the number of cyber attacks[1] every day. Hackers are perturbing the daily activities of users with multiple types of attacks which begin from observing the user activities to intervening in the network and disrupting the entire working of the system.

Malware is a significant cyber-attack[2]. The word malware is a hyponym derived from 'mal' for 'malicious' intent and 'ware' is for 'software'. So, it is a software or program written with malicious intent. Although antiviruses[3] are built with the ability to detect and remove malware with the existing signatures of malware. However, they lack the accuracy in detecting new and unknown malware whose signatures are not found in the antivirus database. These unknown malwares are intelligently coded to change their form and behavior thereby they are undetected by the antiviruses and sandboxes to deceive the detection. Obfuscated malware is metamorphic malware that can hide itself from detection. Since 1980, enough research has been done in the field of malware detection, however, still there are major challenges in this field detecting unknown malware, optimizing the detection rate, detecting obfuscated and evasive malware, predicting malware before the attack, identifying the path of malware propagation, mitigating the flow of malware and recovery from malware infection and many.

Our contribution to this paper:
1. Implementing a multi-class classification model for identifying multiple obfuscated malwares to choose the proper course of action for its mitigation.
2. A Random Forest Classifier has been proven to demonstrate impressive accuracy for selecting important features and for both binary and multi-class classification.
3. Compare the proposed model with the existing dilated CNN model intended for detecting obfuscated malware.

Further, the rest of the paper is organized as mentioned here. In section II brief introduction to obfuscated malware is provided, in section III background and related work is discussed, and in section IV and V proposed methodology and implementation details are presented. Discussing the results in section VI we provide concluding remarks in section VII followed by references.

## II. OBFUSCATED MALWARE

Obfuscation is a software engineering strategy to conceal software from its internal structure and functionality. Malware developers are using these techniques to alter the malware features and behavior such that it can be hidden from malware detection systems.

### A. Obfuscation methods:

Ilsun You [4] classifies malware into encrypted, polymorphic, oligomorphic, and metamorphic malware. Encrypted malware is associated with an encryptor and decryptor. It evades detection by encrypting with different keys during infection, thereby generating different signatures and confusing the antivirus scanner or any ML-based detector. Meanwhile, the decryptor recovers the main body of malware when the infected file is run. As there is a feasibility of mutation of decryptor from one generation to another generation, oligomorphic malware with multiple decryptors were devised. Further to complicate the detection polymorphic malware with an infinite number of decryptors were coded. Dead code insertion and other techniques like the usage of mutation engines were employed to generate such malware. Advanced malware is metamorphic with auto mutation techniques to evolve themselves as and when it is propagating in a network.
According to S. Schrittwieser et al., there are three software obfuscation techniques data obfuscation, static code rewriting, dynamic code rewriting[5] in the context of protecting the software. In data obfuscation the program data is split or merged into several blocks, thus preventing the attacker from evading the software. Dynamic code rewriting makes use of packers and encryptors to alter the program behavior during runtime. SubVirt [6] is one such tool used for code virtualization. Unlike dynamic code rewriting, static code rewriting transforms code during compilation with semantic replacement and substitutions. Injecting dead codes and rearranging the basic blocks of control flow would mislead the reverse engineering of software.

Lichen Jia [7] identifies three types of obfuscation methods used by malware viz. binary, source code level, and packed obfuscation methods. Adversarial examples were developed using these obfuscation methods to evaluate learning-based malware detection systems (LB-MDS). With the frequency of each obfuscation method used in its corresponding obfuscation space, there is a decrease in the accuracy of LB-MDS.

*B. Dataset Description*

Detection of malicious processes and programs is revitalized with the application of memory engineering and forensic analysis to capture vital characteristics and behaviors hidden in obfuscated malware. Canadian Institute of Cybersecurity from the University of New Brunswick has assimilated the CIC MalMem 2022 dataset using Memory feature engineering[8]. This malware dataset is composed of features extracted through memory analysis of memory dump processing done in debug mode. Being an updated and balanced dataset, it consists of 2916 samples of benign, 986 Ransomware samples, 982 Spyware and 948 Trojan Horse samples. Each family of malware has 5 subfamilies of data samples. Being a balanced dataset, it is very useful for our research.

*C. Detection of Obfuscated Malware*

Extreme Learning machines [9] are employed to detect obfuscated malware using the CIC MalMem 2022 dataset. Accuracy and geometric mean of sensitivity and specificity are the metrics used for evaluation. Authors have worked on standard, regularized, and unbalanced ELM methods for binary and multiclass classification of obfuscated malware. Extracting the training time and testing time based on the number of neurons and other metrics, it is shown that accuracy increases with the number of neurons with a maximum accuracy superior to 90% for binary classification but not for multiclass classification.

Dilated CNN model is employed in the classification of obfuscated malware [10]. Its architecture consists of 4 blocks with two convolutional layers, a dropout layer, and a batch normalization layer. For binary classification, sigmoid activation function and binary cross entropy loss function are used. For classifying multiple malware, one hot encoder and Softmax activation function are used. Focal loss function is applied to deal with imbalanced dataset issues. They achieved 99.92% accuracy with Adam Optimizer and for 100 epochs. But, 81.83% accuracy in classifying multiple malware even with 500 epochs.

Another similar experiment was conducted for the detection of Obfuscated malware using an Artificial Neural Network[11]. With three hidden layers of the neural network, activation function ReLU for the hidden layer and Softmax for the outer layer, the number of nodes used was 64 for the first and second hidden layers which subsequently doubled in the third layer to 128. With a batch size of 1024 and 100 epochs, this model identified malware with 99.72% accuracy. As the number of epochs and training time increased the loss was decreased and almost tending to zero. There was no attempt made to classify multiple families of malware in this work.

With a similar dataset Random forest algorithm is used to detect obfuscated malware in the cloud environment[12] which is preceded by the application of nature-inspired optimization techniques for feature selection, Viz. Cuckoo Search Algorithm(CSA), wrapper-based Binary Bat Algorithm(BBA), Particle Swarm Optimization(PSO), and Mayfly algorithm(MA). Although these algorithms decrease the selection of feature set, however, improve the classification accuracy. The model achieved an accuracy of 99.99% with {MA, RF}, 99.91% with {PSO, KNN} and 99.10% with {PSO, SVM} for binary classification, though PSO is excellent for feature selection, multiple malware detection is not addressed.

III. BACKGROUND AND RELATED WORK

*A. Role of AI in generating Obfuscated Malware*

AI techniques are employed in [13] for preparing obfuscated malware by inserting NOP instructions via deep reinforcement learning. It is apparent that, machine learning models used in malware detection systems can be fooled by adversarial examples. Convolutional Neural Network is implemented to insert dead codes at optimal positions, thereby the resulting executable gets a mislabel from the machine learning classifier.

Obfuscated malware generated by [14] using adversarial deep reinforcement learning, employing an efficient action control strategy for generating new malware to defend against LB-MDS. It has been experimentally proved that 67% of the malware generated by this model is efficient in escaping from detection. The new metamorphic malware generated by this model possesses uplifted misclassification and enhanced evasion probability.

Prominent API features of 11 families of malware are extracted from the Cuckoo sandbox by [15]. To represent the extracted features, A feature extraction algorithm, and procedures for feature reduction and representation are proposed. KNN, RF, and DT multiclass classifiers are used to classify 11 families of malware with a high training accuracy of 95.7% but testing accuracy is not highlighted. Although it is found to be time-consuming to extract the dynamic features from the API call traces, overcoming which, API call sequence analysis serves as a major feature of analysis for Obfuscated malware detection.

*B. Random Forest*

Random Forest[16] is a versatile supervised machine learning algorithm for both classification and regression tasks. It is powerful by growing multiple decision trees and aggregating the results of multiple decision trees for better decision-making. It is successful in giving accurate and stable results for various complex problems beginning from image classification, and image segmentation, to cancer cell detection. Also, it has the capability of adaptability to extend its application to multidimensional problems. Various versions of Random Forest like Multinomial random forest, Oblique Random Forest, Random Credal Random Forests etc., are successfully implemented and built in the python libraries.

*C. Target Multiple malware*

We intend to detect three types of obfuscated malware from our trained model viz. ransomware, trojan, and spyware. Ransomware is a type of malware that encrypts the files on the disk to demand a ransom from the victim but without a guarantee that paying the ransom will fetch the access back. Problems pertaining to ransomware are growing rapidly because of the obfuscation techniques adopted by the malware developers. Spyware is a type of malware that performs passive attacks by recording user behavior and activities to transfer third-party networks, more dangerous than active attacks. Trojan is another malware executing malicious activities in the background with the disguise as harmless programs. Trojan Horse by the name 'Animal' appeared in 1974, executed without authorization to copy or replicate to every directory in a user system, it can execute endless activities in the background.

## IV. PROPOSED METHODOLOGY

In this work, we propose a machine learning model which is a result of experimental analysis to detect obfuscated malware and to identify the class of the obfuscated malware. Here we use the CIC MalMem 2022 Dataset as mentioned in Section II B, to identify the class of a new sample of malware as spyware, trojan or ransomware. We have evaluated State of the Art (SoTA) binary classifiers and multi-class classifiers with CIC Malmem 2022 dataset to derive the metrics for comparison and further analysis.

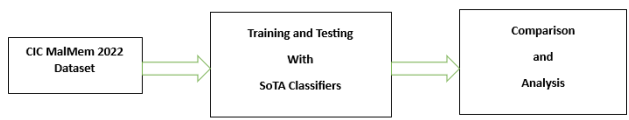

Fig. 1. Actual Workflow of the Model

We conducted two experiments with State of the Art (SoTA) models in Machine Learning for Binary classification and Multi-class Classification. In Binary Classification, a given sample can be identified as malware or non-malware. As per our observation from the literature survey, an enormous amount of work has been implemented in such classification. But, identifying the new sample as malware or benign, does not provide a proper insight on the specific type of malware attack and the course of action to be taken to mitigate the propagation of such malware. Because of the variants of malware like ransomware, spyware, trojan, backdoors, rootkits, viruses, etc., identifying the type of the malware will be helpful for suitable action to be taken to stop and/or recover the adverse effect caused by the malware, which will further aid in mitigating the progress of the malware as well as recovering from the loss caused by the malware in a system or a network. This would also aid in identifying the source of the attack.

In the first experiment of Binary classification, the SoTA classifiers considered are Logistic Regression, Naïve Bayes Classification, Linear SVM classification, Decision Tree, and Random Forest classifiers. Similarly, The SoTA models considered for Multi-class Classification are Naïve Bayes Classification, Decision Tree, Random Forest, Gradient Boosting, and K-Nearest Neighbor. With this intersection, SoTA models can also be studied for their application in such problems.

| Binary classifier Implemented | Multi-class classifier Implemented |
|---|---|
| Logistic Regression | Naïve Bayes Classification |
| Linear SVM classification | Decision Tree |
| Naïve Bayes Classification | Random Forest |
| Decision Tree | Gradient Boosting |
| Random Forest | K-Nearest Neighbor |

Table 1: List of SoTA Classifier models considered.

Consuming updated datasets for cyber security models plays a major role in enhancing the performance, hence we use CIC MalMem2022 dataset which has 58,596 samples. Out of which 80% is taken for training and the remaining 20% is reserved for testing. A baseline of SoTA algorithms is implemented in Python by taking libraries from the scikit learn toolkit. Hyperparameters are optimally chosen by performing rigorous random searches. After performing several hundred iterations optimal hyperparameters were finalized. Before the commencement of the actual experiment, feature engineering is performed to select the relevant features.

Cleansing the data is the initial step in feature engineering, wherein, specified labels of rows and columns are dropped. Especially, when dealing with multi-index labels on different levels this can be achieved by specifying the corresponding axis, index, column names, or levels, thus specific labels from rows and columns can be removed from the data frame without affecting the original data frame, unless required. Further categorization is done to convert categorical variables to indicator variables for powerful representation in statistical modeling for machine learning. To handle categorical variables, one hot encoding is employed, which provides accurate options for controlling prefixes, and suffixes by handling missing values. Whenever data distribution is not Gaussian, ensuring the values within the range will equivalently contribute to the data analysis. MinMax scaler is used to transform data by scaling the features within the range without affecting the shape of the original data distribution. Finally splitting the dataset into 80:20 completes the feature engineering step.

## V. IMPLEMENTATION

Dividing the dataset into training and testing in 80:20 ratio, the binary and multiclass classifier models are implemented using Python and Scikit learn libraries. After the training, predicting a label of a new sample is executed which returns the learned label from the object in the array. This is followed by deriving the metrics from the prediction. For multi-class classification, we employ the Adam optimizer with a Sigmoid activation function and sparse categorical cross-entropy loss function.

## A. Binary Classification of Malware

With reference to, Fig. 2, $M_1, M_2, \ldots M_n$ represents the machine learning classifiers implemented from scikit library. The dataset has a sample $X$ belongs to $\{X_1, X_2 \ldots X_n\}$ with features $F\{F_1, F_2 \ldots F_n\}$ defining a mapping $X \rightarrow F$. Identifying the class $Y$ of $X$ is the major objective of this experiment. $Y$ can be 0 or 1 for benign and malware. $A_1, A_2, \ldots A_n$ are the accuracies derived from the models $M_1, M_2, \ldots M_n$. Comparing these accuracies, we evaluate and analyze the outstanding performer among all the binary classifiers. With a similar ground rule, the multi-class classifiers are also analyzed as in Fig. 3 for which Y can be 0,1,2 and 3 for benign, spyware, ransomware, and trojan.

With this major objective, we carried out the experiment to create a baseline of five machine-learning algorithms. As mentioned earlier, implementation is undertaken with the scikit learn library. Basic and non-parameterized functions were used for Logistic Regression and Naïve Bayes classifier. For Decision Tree minimum samples of leaf used are 3 with a maximum depth of 10, entropy as the criteria, and log2 max features are used. Repeating the same parameters for Random Forest with a number of estimators as 30. $C = 1$ was the right choice for Linear SVM.

## B. Multi-class Classification of Malware

The major objective of this experiment is to create a baseline of five machine learning algorithms. Like the binary classification experiment described in section V A, the implementation is made with the scikit learn library. Basic and non-parameterized functions were used for Naïve Baye's classifier. For the Decision Tree, the minimum samples of leaf used are 16 with a maximum depth of 12, entropy as the criteria, and log2 max features are used. Repeating the same parameters for Random Forest with number of estimators as 30, min_samples split =4 and max depth as 40. With the batch size of 2000 and just 10 epochs we achieved better accuracy with Random Forest. Learning rate of 0.2 was the right choice for Gradient Boosting. The ML models are tested and evaluated using the following metrics.

i) Accuracy: metric used to measure the correctness in the classification. Ratio of samples identified correctly to the total samples.

$$Accuracy = \frac{No. \, of \, samples \, correctly \, identified}{Total \, samples}$$

ii) Precision: metric used to measure the preciseness of the model in predicting positive samples.

$$Precision = \frac{positive \, samples \, rightly \, predicted \, as \, positive}{Total \, positive \, samples \, identified}$$

iii) Recall: metric used to measure how many of predicted positive samples are correct.

$$Recall = \frac{positive \, samples \, rightly \, predicted \, as \, positive}{Total \, rightly \, identified \, positive \, samples \, + \, Total \, rightly \, identified \, negative \, samples}$$

iv) F1-score: metric that gives balance factor between precision and recall, its value is directly proportional to the performance.

$$F1 \, score = 2 \, x \, \frac{Precision \, x \, Recall}{Precision \, + \, Recall}$$

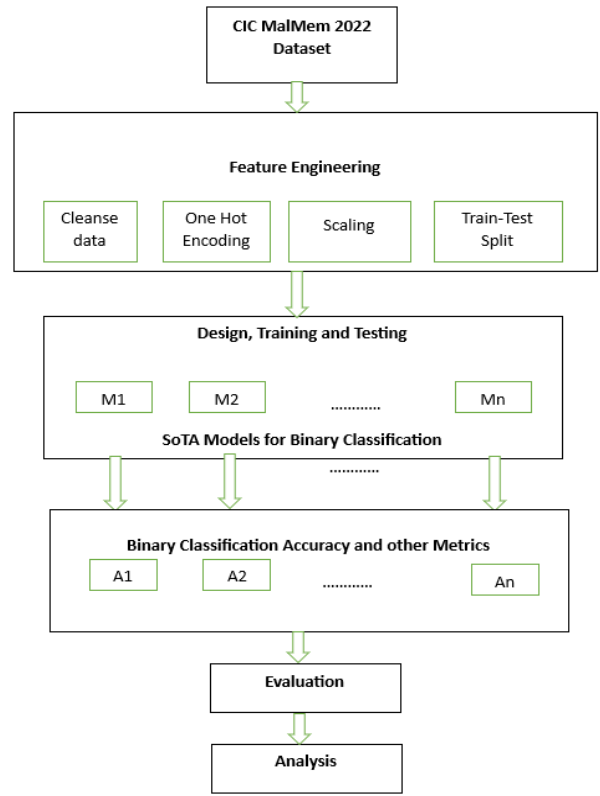

Fig. 2. Binary Classification Model

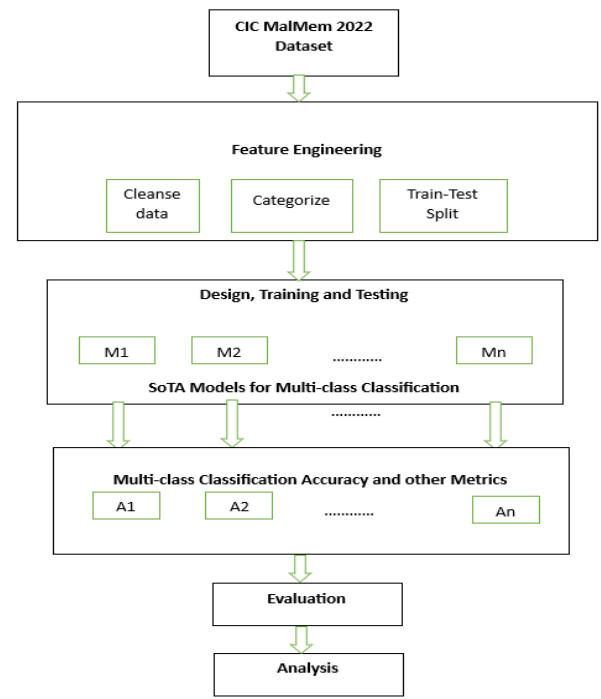

Fig. 3. Multi-class Classification Model

## VI. RESULTS AND DISCUSSION

### A. Results of Binary Classification

The following are the results deduced from the experiment details discussed in the previous section. The values of the metrics accuracy, precision, recall, and F1 score of binary and multi-class classification are tabulated in Tables 3 and 4 respectively.

| Model | Accuracy | Precision | Recall | F1 Score |
|---|---|---|---|---|
| Logistic Regression | 99.56% | 99.42 | 99.71 | 99.56 |
| Linear SVM classification | 99.88% | 99.88 | 99.88 | 99.88 |
| Naïve Bayes Classification | 99.21% | 98.78 | 99.65 | 99.21 |
| Decision Tree | 99.99% | 99.98 | 99.982 | 99.99 |
| **Random Forest** | **99.982%** | **99.982** | **99.982** | **99.982** |
| ANN [11] | **99.72%** | **~100.0** | **99.9** | **~100** |
| MLP Classifier[10] | **99.70%** | **99.70** | **99.70** | **99.70** |
| kNN classifier [10] | **99.96%** | **99.96** | **99.96** | **99.96** |
| Dilated CNN [10] | **99.88%** | **99.88** | **99.88** | **99.88** |

Table 2: Binary Classification-Results.

It is evident that for the hyperparameters chosen by our model, all models are performing equivalent but Random Forest performance is consistent in all the tests. Results of ANN, MLP Classifier, KNN Classifier, and dilated CNN which are existing models are also shown here for comparison.

*B. Results of Multi-class Classification*

| Model | Accuracy | Precision | Recall | F1 Score |
|---|---|---|---|---|
| Naïve Bayes Classifier | 68.86% | 68.86 | 73.26 | 64.51 |
| Decision Tree | 84.67% | 84.89 | 84.92 | 84.90 |
| **Random Forest** | **89.07%** | **87.63** | **87.62** | **87.62** |
| Gradient Boosting | 83.84% | 83.84 | 83.84 | 83.83 |
| K-Nearest Neighbor | 79.80% | 79.80 | 79.85 | 79.81 |
| Dilated CNN [10] | 81.83% | 72.71 | 72.72 | 72.71 |

Table 3: Multi-class Classification-Results.

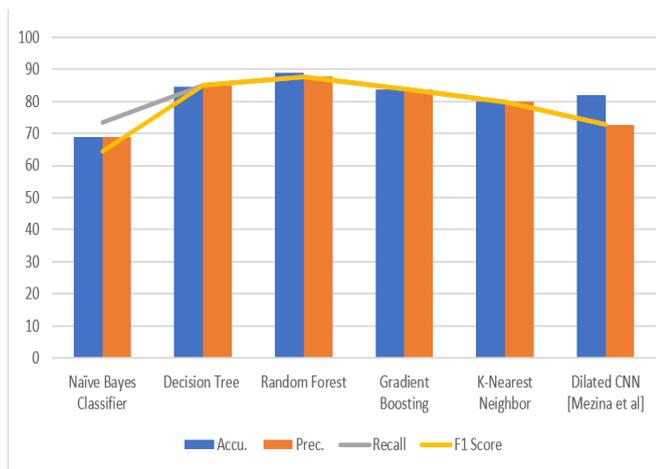

Fig. 4. Multi-class Classifiers - Results

It is evident from the results of multiclass classification that Decision Tree and Gradient Boosting are closer in performance, but Random Forest is performing outstanding among other models, although 89% is not superior, but it is comparatively better result obtained so far. A combo chart in Fig.4 shows the distribution of metrics. Comparing these results with Anzhelika's Dilated CNN model, our proposed model is better with +8%. A similar experiment was carried out by Lamia Pervan using ANN, although binary classification results were best, the model performance decreased for multiclass classification. The confusion matrices derived from our experiment are as shown in Fig.5.

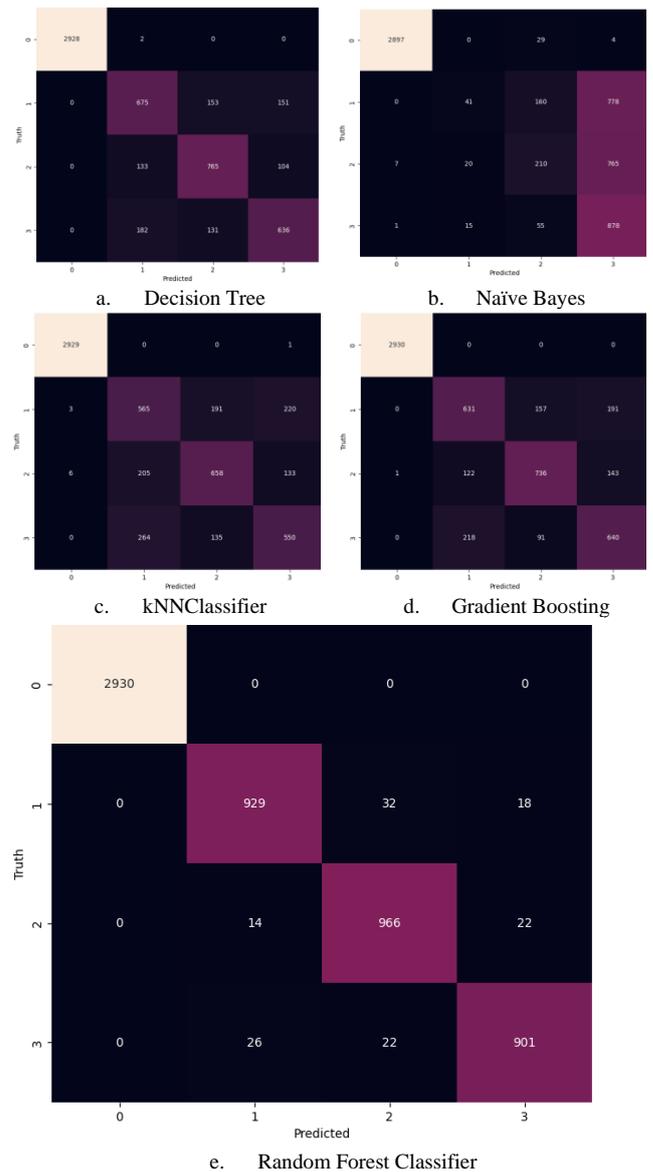

a. Decision Tree  b. Naïve Bayes

c. kNNClassifier  d. Gradient Boosting

e. Random Forest Classifier

Fig. 5. Confusion matrices with Multi-class Classifiers

VII. CONCLUSION

Obfuscated malware detection is one of the hot topics of research in the field of AI-infused cyber security. Although a good amount of work is found in classifying a sample as malware or non-malware, to the best of our knowledge significant research is lacking in detecting multiple malware in a single model. In this paper, we have implemented a Machine Learning-based cybersecurity model for multi-class classification of obfuscated malware to detect three types of malware viz. spyware, ransomware, and trojan. We have compared the results of our work with existing works and presented that our proposed model performance is 8% better than the existing models with the better hyperparameters we chose. With the Random Forest algorithm and considerable hyperparameter tuning, we achieved an accuracy of 89.07% in classifying multiple obfuscated malware. Although there is further scope for improvement in achieving still higher accuracy, extensive experiments are being conducted for further improvement in accuracy.

## REFERENCES

[1] H. S. Berry, "The Evolution of Cryptocurrency and Cyber Attacks," in *2022 International Conference on Computer and Applications (ICCA)*, Dec. 2022, pp. 1–7. doi: 10.1109/ICCA56443.2022.10039632.

[2] B. Marais, T. Quertier, and S. Morucci, "Malware and Ransomware Detection Models," pp. 1–8, 2022, doi: https://doi.org/10.48550/arXiv.2207.02108.

[3] M. Botacin *et al.*, "AntiViruses under the microscope: A hands-on perspective," *Comput. Secur.*, vol. 112, p. 102500, Jan. 2022, doi: 10.1016/j.cose.2021.102500.

[4] I. You and K. Yim, "Malware Obfuscation Techniques: A Brief Survey," in *2010 International Conference on Broadband, Wireless Computing, Communication and Applications*, Nov. 2010, pp. 297–300. doi: 10.1109/BWCCA.2010.85.

[5] S. Schrittwieser, S. Katzenbeisser, J. Kinder, G. Merzdovnik, and E. Weippl, "Protecting Software through Obfuscation," *ACM Comput. Surv.*, vol. 49, no. 1, pp. 1–37, Mar. 2017, doi: 10.1145/2886012.

[6] S. T. King and P. M. Chen, "SubVirt: implementing malware with virtual machines," in *2006 IEEE Symposium on Security and Privacy (S&P'06)*, 2006, pp. 14 pp. – 327. doi: 10.1109/SP.2006.38.

[7] L. Jia, Y. Yang, B. Tang, and Z. Jiang, "ERMDS: A obfuscation dataset for evaluating robustness of learning-based malware detection system," *BenchCouncil Trans. Benchmarks, Stand. Eval.*, vol. 3, no. 1, p. 100106, Feb. 2023, doi: 10.1016/j.tbench.2023.100106.

[8] T. Carrier, P. Victor, A. Tekeoglu, and A. Lashkari, "Detecting Obfuscated Malware using Memory Feature Engineering," in *Proceedings of the 8th International Conference on Information Systems Security and Privacy*, 2022, pp. 177–188. doi: 10.5220/0010908200003120.

[9] L. Igor Moraga, J. P. R. Malcó, D. Zabala-Blanco, R. Ahumada-García, C. A. Azurdia-Meza, and A. D. Firoozabadi, "Detection of Obfuscated Malware by Engineering Memory Functions Applying ELM," in *2023 IEEE Colombian Conference on Applications of Computational Intelligence (ColCACI)*, Jul. 2023, pp.1–6. doi:10.1109/ColCACI59285.2023.10226058.

[10] A. Mezina and R. Burget, "Obfuscated malware detection using dilated convolutional network," in *2022 14th International Congress on Ultra Modern Telecommunications and Control Systems and Workshops (ICUMT)*, Oct. 2022, pp. 110–115. doi: 10.1109/ICUMT57764.2022.9943443.

[11] L. P. Khan, "Obfuscated Malware Detection Using Artificial Neural Network (ANN)," in *2023 Fifth International Conference on Electrical, Computer and Communication Technologies (ICECCT)*, Feb. 2023, pp. 1–5. doi: 10.1109/ICECCT56650.2023.10179639.

[12] M. R. Ghazi and N. S. Raghava, "Machine Learning Based Obfuscated Malware Detection in the Cloud Environment with Nature-Inspired Feature Selection," in *2022 5th International Conference on Multimedia, Signal Processing and Communication Technologies (IMPACT)*, Nov. 2022, pp. 1–5. doi: 10.1109/IMPACT55510.2022.10029271.

[13] D. Gibert, M. Fredrikson, C. Mateu, J. Planes, and Q. Le, "Enhancing the insertion of NOP instructions to obfuscate malware via deep reinforcement learning," *Comput. Secur.*, vol. 113, p. 102543, Feb. 2022, doi: 10.1016/j.cose.2021.102543.

[14] M. Sewak, S. K. Sahay, and H. Rathore, "DOOM: A novel adversarial-DRL-based op-code level metamorphic malware obfuscator for the enhancement of IDS," *UbiComp/ISWC 2020 Adjun. - Proc. 2020 ACM Int. Jt. Conf. Pervasive Ubiquitous Comput. Proc. 2020 ACM Int. Symp. Wearable Comput.*, pp. 131–134, 2020, doi: 10.1145/3410530.3414411.

[15] C. C. San, M. M. S. Thwin, and N. L. Htun, "Malicious Software Family Classification using Machine Learning Multi-class Classifiers," 2019, pp. 423–433. doi: 10.1007/978-981-13-2622-6_41.

[16] L. Brieman, "Random Forests," *Machine. Learning.*, vol. 45, no. Oct-2001, pp. 5–32, 2001, doi: https://doi.org/10.1023/A:1010933404324.